\documentclass[12pt, a4paper]{iopart}

\usepackage[]{graphicx}  
\usepackage{wrapfig}
\usepackage{courier}
\usepackage{multirow}
\usepackage{cite}
\usepackage[rightcaption]{sidecap}

\usepackage{xcolor}

\begin{document}

\title[]{Impact of magnetic islands on plasma flow and turbulence in W7-X}

\author[1]{T. Estrada$^1$, E. Maragkoudakis$^1$, D. Carralero$^1$, T. Windisch$^2$, J.L Velasco$^1$, C. Killer$^2$, T. Andreeva$^2$, J. Geiger$^2$, A. Dinklage$^2$, A. Kr\"amer-Flecken$^3$, G. A. Wurden$^4$, M. Beurskens$^2$, S. Bozhenkov$^2$, H. Damm$^2$, G. Fuchert$^2$, E. Pasch$^2$ and the W7-X team$^2$}

\address{$^1$ Laboratorio Nacional de Fusi\'on. CIEMAT, 28040 Madrid, Spain}
\address{$^2$ Max-Plank-Institut f\"ur Plasmaphysik, D-17491 Greifswald, Germany}
\address{$^3$ Institut f\"ur Energie- und Klimaforschung - Plasma Physik, Forschungszentrum J\"ulich, 52425 J\"ulich, Germany}
\address{$^4$ Los Alamos National Laboratory, PO Box 1663, Los Alamos, NM 87545 USA}

\ead{teresa.estrada@ciemat.es}


\begin{abstract}
{

The effect of magnetic islands on plasma flow and turbulence has been experimentally investigated in the stellarator W7-X. Magnetic configurations with the 5/5 magnetic island positioned at the plasma edge, inside the last closed flux surface, are studied. The main diagnostic used in the present work is a V-band Doppler reflectometer that allows the measurement of the perpendicular plasma flow and density fluctuations with good spatial resolution. 
A characteristic signature of the 5/5 magnetic island  is clearly detected in the perpendicular flow profile. The comparison of the experimental flow and the neoclassically driven $E\times B$ flow indicates that the island contribution to the flow is maximum at the island boundaries and close to zero at the island O-point. Besides, a reduction in the density fluctuation level is found nearby the island O-point.
The similarities between these observations and those found in other devices and in gyrokinetic simulations are discussed.
}
\end{abstract}

\section{Introduction}

The effect of magnetic islands on plasma confinement and transport has been studied during the last decades in different fusion devices~\cite{Erckmann:1985,Renner:1989,LopesCardozo:1997,Shaing:2001,Brakel:2002,Ida:2001,Harris:2004,Ascasibar:2008}. 
Magnetic islands placed in the confinement region may cause confinement degradation in the absence of magnetic shear~\cite{Erckmann:1985,Renner:1989,Brakel:2002,Ascasibar:2008}; with finite magnetic shear however, magnetic islands have been found to ease the formation of internal transport barriers both, in tokamaks and in helical devices~\cite{Joffrin:2003,Austin:2006,Estrada:2004,Castejon:2004,Estrada:2007,Narushima:2011}, and the transition to High confinement mode (H-mode) in helical devices~\cite{Hirsch:2008,Sano:2005,Estrada:2009,Estrada:2011b}. 
These results have been interpreted in terms of local changes in the perpendicular flow velocity associated to magnetic islands which may result in a reduction of plasma turbulence~\cite{Hidalgo:2000,Garcia:2001}.

Pioneering experiments carried out in LHD with a static 1/1 magnetic island produced by external perturbation coils have shown how the poloidal flow velocity behaves inside the magnetic island and how it depends on the island width~\cite{Ida:2001}. The flow shows a perturbation that is radially symmetric across the magnetic island in cases of small islands but becomes asymmetric and the flow sign reverses at the O-point creating a vortex-like structure when the island width is large.
Similar experiments have been performed in J-TEXT~\cite{Zhao:2015} and in KSTAR~\cite{Choi:2017}, where the multi-scale interaction between plasma flow and fluctuations were measured nearby static magnetic islands driven by resonant magnetic perturbations. The plasma flow increases at the island boundaries while the density fluctuations drop inside the magnetic island and increase at the island boundaries. Besides, in J-TEXT~\cite{Zhao:2015}, the flow is enhanced at the outer island boundary when the magnetic island approaches the Last Closed Flux Surface (LCFS).
In TJ-II, the effect of magnetic islands on perpendicular plasma flow and turbulence has been experimentally investigated in dynamic magnetic configuration scans~\cite{Bondarenko:2010,Estrada:2016}. 
As in previous experiments, the perturbation in the flow changes with the island size and has an impact in the density fluctuations inside the magnetic island. In DIII-D, a reduction in the density fluctuations associated to large magnetic islands has been also measured in the plasma core~\cite{Bardoczi:2016}.
Differences in the perpendicular flow measured across the island O-point and the island X-point have been found in HL-2A~\cite{Jiang:2017}, with stronger flow-shear at the island boundaries at the O-point and a nearly flat flow profile at the X-point. This difference has been also studied theoretically~\cite{Hahm:2021} showing that the flow-shearing near the X-point is important for the turbulence penetration into the island. Such a turbulence spreading into the island has been demonstrated experimentally in several devices: DIII-D~\cite{Ida:2018}, HL-2A~\cite{Jiang:2019}, and KSTAR~\cite{Choi:2021}.
Finally, gyrokinetic simulations have been also performed to address this topic~\cite{Banon_Navarro:2017}. The effect of a static magnetic island on plasma flows, turbulence and transport has been studied as well as the scaling of these effects with the island width. Besides, the effect of a radial asymmetry of the magnetic island on flows and turbulence is described. The simulations show similarities with the experimental findings and can be considered as a guideline to interpret the latter.

The optimised, superconducting stellarator W7-X offers the possibility to explore different magnetic configurations~\cite{Geiger:2014}.
Depending on the magnetic configuration, natural magnetic islands form at the plasma boundary which are intersected with the divertor target plates forming an island-divertor with X-points  running around helically~\cite{Klinger:2019}.
In general, island divertor operation is possible in configurations with edge rotational transform equal to 5/m with $m= 4, 5$ or $6$.
In the present experiments, however, limiter magnetic configurations with edge rotational transform slightly above one are studied.
In these configurations the 5/5 magnetic island chain forms at the plasma edge inside the LCFS. Details on the technical realization and on the confinement and equilibrium properties of the magnetic configuration scans are reported in~\cite{Andreeva:2019,Andreeva:2021,Geiger:2021}. Besides, island localized MHD-activity is described in~\cite{Andreeva:2019,Wurden:2019,Han:2021}.
The present work reports radially-resolved measurements of perpendicular plasma flow and density fluctuations by Doppler reflectometry (DR) and discuss their interaction across the 5/5 magnetic island at the plasma edge of W7-X.
These results extent previous knowledge on the link between magnetic islands and transport barriers through the formation of sheared-flow layers with reduced turbulence, and may also help in the understanding of the influence of magnetic islands on the Scrape-Off Layer (SOL) transport and eventually on the divertor properties~\cite{Killer:2019,Hammond:2019}.

The remainder of the paper is organised as follows. The experimental set-up is described in section 2 and the experimental results are shown in section 3. Finally, the summary and discussion are included in section 4.

\section{Experimental set-up}

W7-X is an optimised superconducting stellarator with major radius $R = 5.5$ m, minor radius $a = 0.5$ m and magnetic field on axis $B_0 = 2.5$ T~\cite{Klinger:2016,Wolf:2017}.
W7-X, being equipped with 50 non-planar and 20 planar superconducting coils, offers the possibility to explore different magnetic configurations~\cite{Geiger:2014}. 
In general, W7-X operates with an island divertor compatible magnetic field structure which is possible in configurations with edge rotational transform equal to 5/6 (low iota), 5/5 (standard) and 5/4 (high iota configuration)~\cite{Klinger:2019}. 
In this work, however, limiter configurations with intermediate rotational transform values between the high iota and the standard configurations are studied. They are called limiter configurations because the LCFS is determined by the intersection of field lines of nested flux surfaces by the target plates of the divertor which acts as a limiter.  In these cases, the 5/5 magnetic island is located at the plasma edge inside the LCFS. 
The plasmas are created and heated by ECH $2^{nd}$ harmonic at 140 GHz. The ECH system~\cite{Erckmann:2007} consists of 10 long-pulse gyrotrons with a power per gyrotron of up to 0.8 MW.

The main experimental results presented in this work have been obtained using Doppler reflectometry (DR). 
A V-band (50-75 GHz) DR system working in O-mode polarization has been used to measure density fluctuations and their perpendicular rotation velocity, $u_\perp$~\cite{Carralero:2020,Estrada:2021}. The reflectometer front end, installed at port AEA21 (toroidal angle $\phi = 72 ^\circ$), uses a single antenna and a set of mirrors for launching and receiving the signal at fixed probing beam angle of $\alpha = 18^ \circ$~\cite{Windisch:2019}. Under these conditions, perpendicular wave-numbers of the turbulence in the range $k_\perp \sim 7-10$ cm$^{-1}$ are measured at the accessible local densities in the range from 2.8 to 6.3 $\times 10^{19}$ m$^{-3}$. 
In the present experiments, the corresponding normalized wave-numbers vary from $k_\perp \rho_i \sim 0.3$ at the plasma edge to 1 at the plasma core.
During each experimental program, the frequency of the reflectometer is scanned in a hopping mode from 50 to 75 GHz, typically in steps of 1 GHz, 10 ms long. Thus, every 250 ms a complete scan is performed. For each probing frequency, the corresponding radial position, $\rho$, and perpendicular wave-number, $k_\perp$, are calculated using the 3D ray-tracing code TRAVIS~\cite{Marushchenko:2014} with the density profile measured by the Thomson Scattering diagnostic~\cite{Pasch:2016} and the magnetic configuration provided by a VMEC-equilibrium calculation. VMEC assumes nested flux surfaces and is therefore no able to properly reflect the existence of the internal 5/5 magnetic islands. However, as shown below, this limitation is properly considered in our evaluation and interpretation of the data.
To estimate uncertainties, a bundle of rays are considered to reflect the trajectories of the $1/e$-amplitude of the DR probing beam. The radial positions of reflection and local incident wave numbers of these rays are used to estimate errors in $\rho$ and $k_\perp$.

The perpendicular rotation velocity of the plasma turbulence measured by DR is a composition of both the plasma $E \times B$ velocity and the intrinsic phase velocity of the density fluctuations: $u_\perp = v_{E \times B} + v_{ph} $. In cases in which the condition $v_{E \times B} \gg v_{ph}$ holds, $E_r$ can be obtained directly from the perpendicular rotation velocity as $E_r = u_\perp \cdot B$. In the present experiments, only the $u_\perp$ profiles are discussed.
Regarding the density fluctuations, the power of the back-scattered DR signal, $S$, is the relevant quantity proportional to $| \delta n|^2$.
It has to be noted that, in general, a microwave generator working with variable frequency produces a different power output at each frequency. Besides, the transmitted power through the transmission line may also depend on the frequency. Therefore, for a proper comparison of the fluctuations measured at different frequencies, a power calibration of the Doppler reflectometer is indispensable~\cite{Estrada:2021}.

\section{Experimental results}

Magnetic configuration scan experiments have been performed in W7-X during the 2018 experimental campaign to study the impact of the rotational transform on plasma confinement~\cite{Andreeva:2019,Andreeva:2021}.  An increase of the plasma energy content and confinement time was found at intermediate limiter configurations between the high iota and the standard magnetic configurations. These intermediate limiter configurations have the 5/5 magnetic island located at the plasma edge nearby $\rho \sim 0.6-0.8$~\cite{Andreeva:2021}. 
During these experiments, DR measurements have been carried out to characterise the perpendicular plasma flow and density fluctuations.

\subsection{Magnetic configuration scan} \label{Magnetic configuration scan}

\begin{figure}[h]
 \center
\includegraphics[width=0.32\columnwidth,trim= 0 0 0 0]{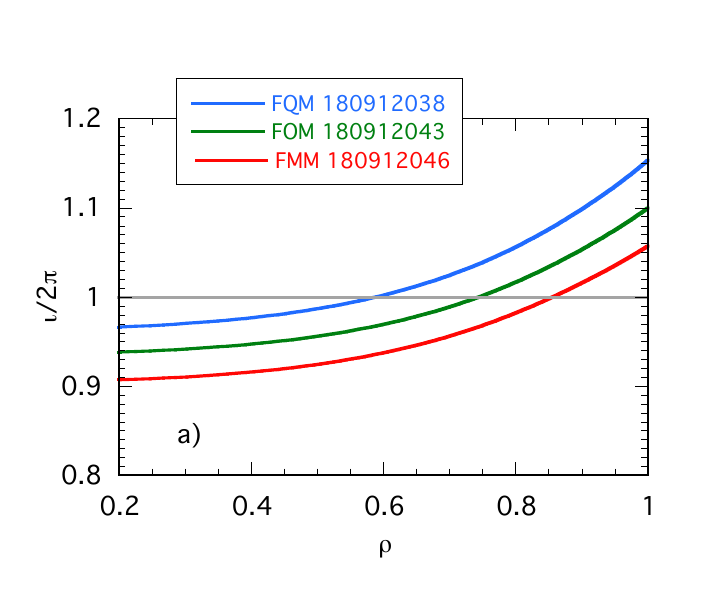}     
\includegraphics[width=0.32\columnwidth,trim= 0 0 0 0]{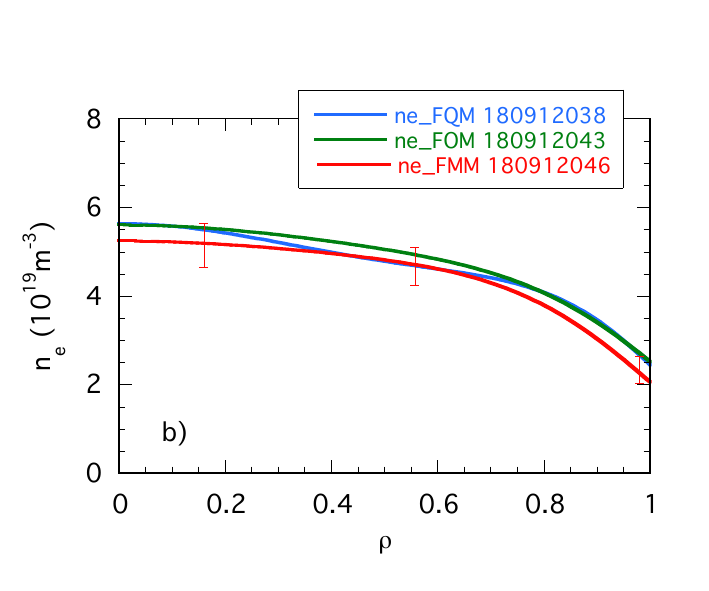}   
\includegraphics[width=0.32\columnwidth,trim= 0 0 0 0]{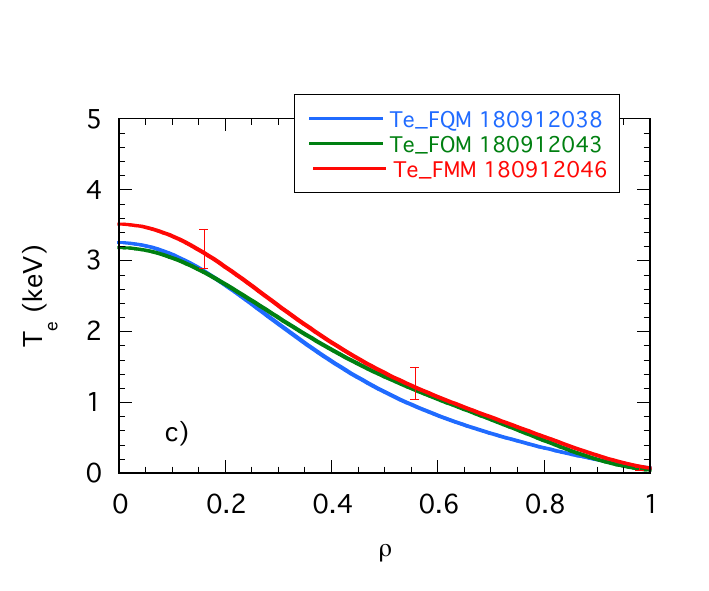}   
 \caption{Vacuum rotational transform profiles of three magnetic configurations: FQM, FOM and FMM (a), and the corresponding electron density and temperature profiles (b \& c) measured by the Thomson scattering diagnostic; these profiles result from the fit to the experimental values whose dispersion is indicated by some representative error bars.}
\label{f:fig_1}
\end{figure}

Three magnetic configurations are explored with edge rotational transform equal to 1.15, 1.10 and 1.05, named FQM001, FOM003 and FMM002, respectively. 
For simplicity, however, in the remainder of the article the configuration type names FQM, FOM, and FMM will be used to refer to these particular configurations.
In these experiments, plasmas are heated with $P_{ECH}=4$ MW and the line integrated density is kept constant along each experimental program at values within the range $n_e \sim 6.5 - 7.0$ $10^{19}$ m$^{-2}$. 
The rotational transform profiles of the three magnetic configurations and the corresponding electron density and temperature profiles measured by the Thomson scattering diagnostic~\cite{Pasch:2016} are shown in figure \ref{f:fig_1}. No pronounced differences are found in the electron density and temperature profiles when the three magnetic configurations are compared. 
Note, that no clear flattening associated to the magnetic island is observed neither in density nor in temperature profiles. This is expected as the Thomson scattering diagnostic line of sight crosses the 5/5 magnetic island near the X-point~\cite{Pasch:2016}.  Thus, a flattening in the pressure profiles, as expected at the O-point if the parallel transport inside the island dominates over the cross-field transport~\cite{Banon_Navarro:2017}, cannot be completely ruled out.

\begin{figure}[h]
 \center 
\includegraphics[width=0.55\columnwidth,trim= 20 20 0 0]{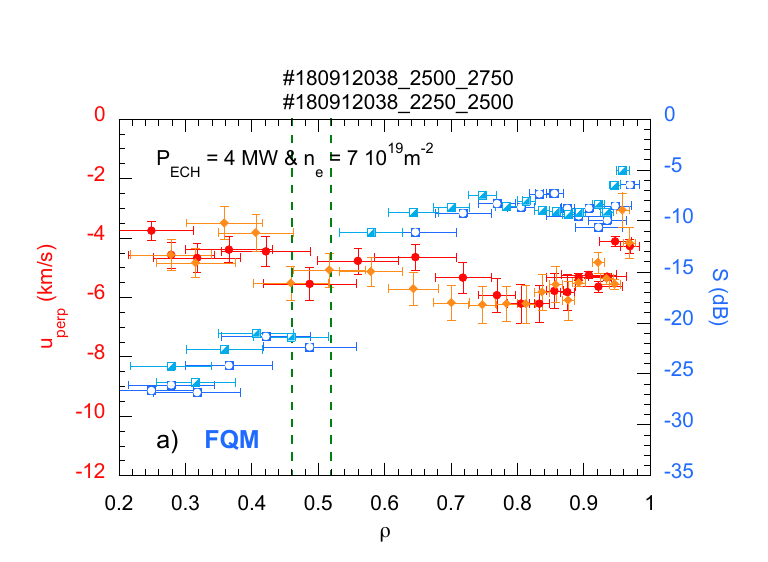}   
\includegraphics[width=0.55\columnwidth,trim= 20 20 0 0]{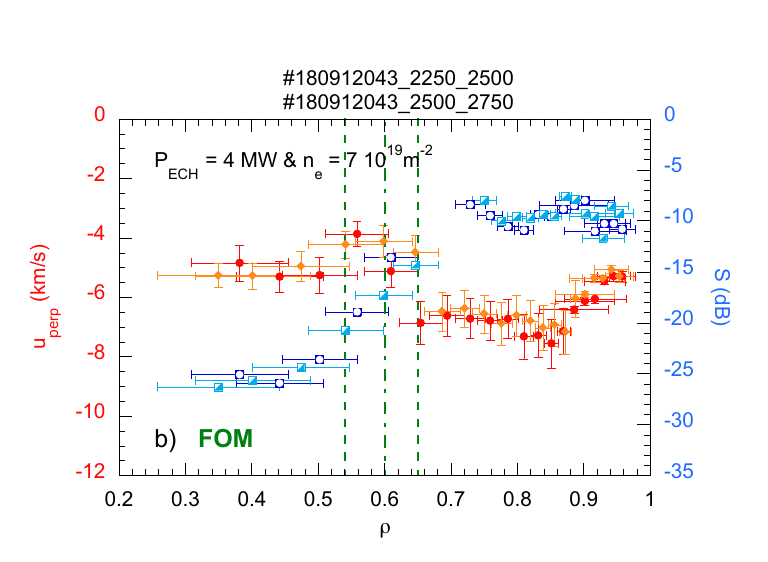}   
\includegraphics[width=0.55\columnwidth,trim= 20 20 0 0]{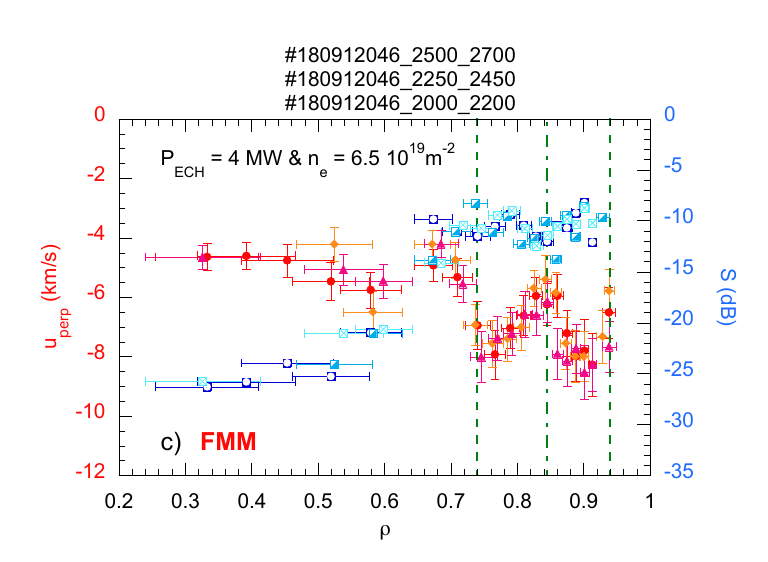}    
 \caption{Radial profiles of perpendicular plasma flow (in reddish colors) and density fluctuations (in blueish colors) measured by DR in three magnetic configurations: FQM (a), FOM (b) and FMM (c). Different symbols and color variations are used for the different time intervals within each experimental program. Vertical green lines indicate the DR measurement points closer to the island boundaries (dashed lines) and to the island O-point (dashed and dotted line) as obtained from the Poincar\'e plots. Refer to the text for more details.}
\label{f:fig_2}
\end{figure}

The DR results obtained in the three configurations are shown in figure  \ref{f:fig_2}. 
The radial profiles of the perpendicular plasma flow (in red) and density fluctuations (in blue), measured in the three magnetic configurations, are shown in figures \ref{f:fig_2}.a (FQM), \ref{f:fig_2}.b (FOM), and \ref{f:fig_2}.c (FMM).
For each experimental program, the profiles measured at different time intervals are represented (using different symbols) showing the reproducibility of the measurements during the stationary phase of the discharges.
As described in~\cite{Wurden:2019}, in the FMM configuration, the so-called Island Localized Modes (ILMs) appear as crashes in the Rogowski coil signals. Each crash has a time duration of about 1-3 milliseconds and produces small drop of the total plasma energy. DR detects these events as spikes in the spectrogram during which the Doppler peak is disturbed.  Therefore, in the analysis of the DR data presented in this paper, time slices corresponding to the crashes have been thoroughly avoided.

At first sight, the perpendicular flow profiles measured in configurations FQM (figure \ref{f:fig_2}.a)  and FOM (figure \ref{f:fig_2}.b) do not show any marked peculiarity and resemble those measured in usual island divertor configurations~\cite{Carralero:2020}.  
In the FMM configuration, however, a very remarkable pattern appears in the flow profile (figure \ref{f:fig_2}.c) showing a $W$-$shape$ in the radial range $\rho \sim 0.70 - 0.95$.

\begin{figure}[h]
 \center
\includegraphics[width=0.2\columnwidth,trim= 0 0 0 0]{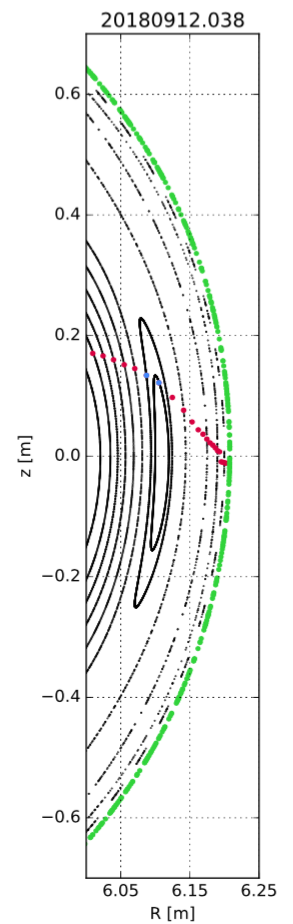} 
\includegraphics[width=0.2\columnwidth,trim= 0 0 0 0]{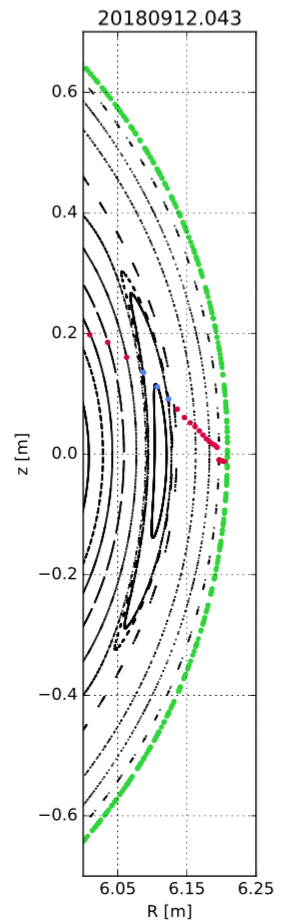}   
\includegraphics[width=0.2\columnwidth,trim= 0 0 0 0]{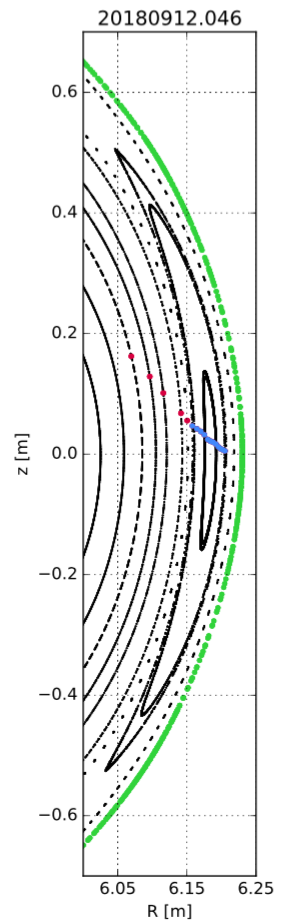}   
 \caption{Poincar\'e plots calculated for the experimental programs shown in figure  \ref{f:fig_2}, in the magnetic configurations FQM (left), FOM (center) and FMM (right). The 5/5 island chain is visible in the three configurations inside the LCFS (shown in green). The red and blue dots represent the DR measurements points.} 
\label{f:fig_3}
\end{figure}

 The $W$-$shape$ pattern in the plasma flow found in the FMM configuration may be associated to the 5/5 magnetic island; this pattern, however, is absent in the FQM and FOM configurations despite the proximity of the three rotational transform profiles. In all three cases, the radial position at which the 5/5 magnetic island is expected falls within the radial range covered by the DR measurements.
In order to clarify this apparent controversial result, the Poincar\'e plots for the three configurations have been obtained from field line tracing calculations~\cite{Bozhenkov:2013}. They are shown in figure \ref{f:fig_3}, where the solid green line denotes the LCFS. The 5/5 island chain is visible inside the LCFS in the three configurations, gradually becoming wider while moving to the plasma boundary from FQM to FMM configurations (from left to right). It is worth noting that it is in the latter, i.e. in FMM, where a better plasma confinement has been found~\cite{Andreeva:2021,Geiger:2021}.

The field line calculations are based on the coil currents for the ideal configuration and take into account the adjustment of the current of the planar coils. This adjustment compensates for the iota change consequence of the deformation of the coil geometry due to the electromagnetic forces of the charged coils~\cite{Andreeva:2021}. 
The possible effect of the plasma current on the magnetic topology has been neglected in the calculations because in these experimental programs the net plasma current stays at very low values, below 1kA, all along the plasma discharges.

In order to estimate the relative position of the DR measurements -calculated assuming nested flux surfaces- with respect to the magnetic island, the DR (x,y,z)-positions are mapped into the Poincar\'e plot.
The (red and blue) dots shown in figure \ref{f:fig_3} represent the DR measurement points. 
This procedure is used as a guideline in the interpretation of the DR measurements, i.e. to know whether the measurements points are crossing the magnetic island and how close to the O-point.
The blue dots represent the DR measurement points located at the 5/5 magnetic island in each magnetic configuration. Clear differences are found when the first two configurations (FQM and FOM) are compared to the third one (FMM). In the first cases only two/three DR probing frequencies measure in the 5/5 island region while in the latter, a large number of DR probing frequencies do, crossing the island much closer to the O-point than in the other two configurations. As a result, the flow structure across the magnetic island can be clearly measured only in the FMM configuration.
From the Poincar\'e plots, the DR measurement points closer to the island boundaries and closer to the island O-point have been identified and have been represented in figure \ref{f:fig_2} as vertical green lines. 
In the FMM configuration (figure \ref{f:fig_2}.c), a rather good matching is found between the island region and the $W$-$shape$ pattern in the perpendicular plasma flow. From the Poincar\'e plot, the estimated island full-width in this case is $\Delta \sim 4$ cm. 
In the FQM and FOM configurations, the few DR probing frequencies measuring in the 5/5 island region show a slight modification in the perpendicular plasma flow (figures \ref{f:fig_2}.a and \ref{f:fig_2}.b), which is almost within the error bars.

\begin{figure}[h]
 \center 
\includegraphics[width=0.48\columnwidth,trim= 20 20 20 0]{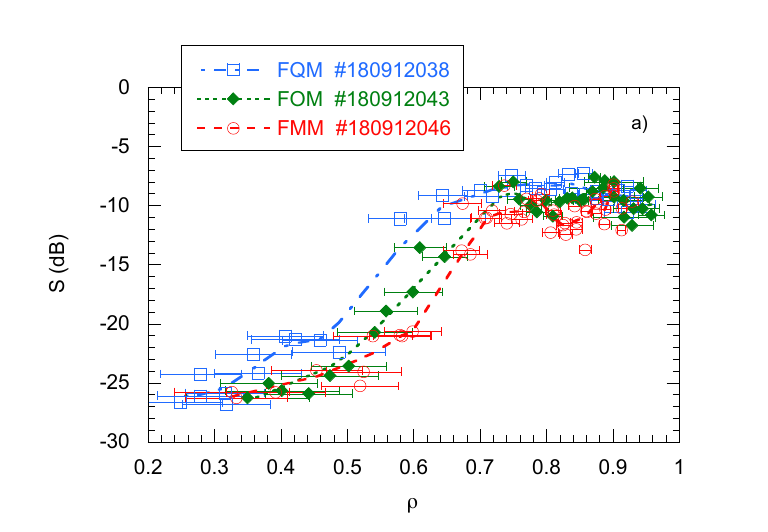}   
\includegraphics[width=0.48\columnwidth,trim= 20 20 20 0]{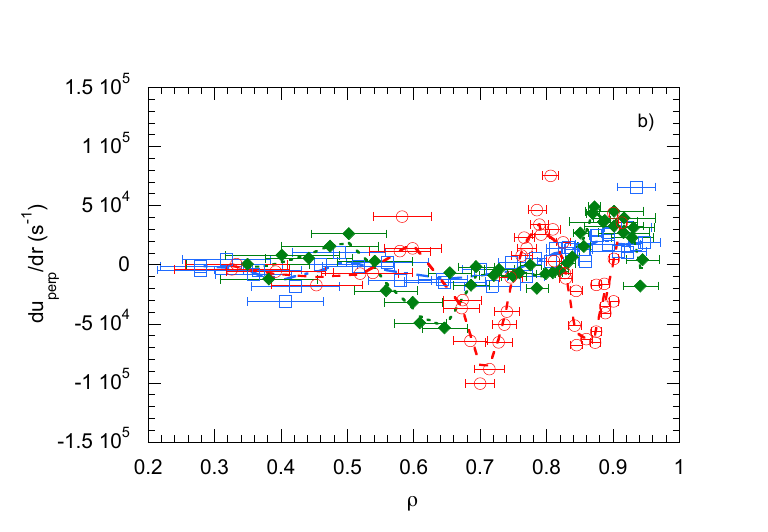}     
 \caption{Radial profiles of density fluctuations (a) and flow-shearing (b) measured by DR in the three magnetic configurations; same experimental programs as shown in figure \ref{f:fig_2}.}
\label{f:fig_4}
\end{figure}

Regarding the density fluctuations (shown in blue in figure  \ref{f:fig_2}), similar profiles are found in the three magnetic configurations. In all three cases, two radial zones can be distinguished, an external region with high fluctuation level, and an internal one towards the plasma core where the fluctuation level drops.  A closer comparison, however, reveals some differences which may be linked to the 5/5 magnetic island. This comparison is shown in figure \ref{f:fig_4}.a, where the density fluctuations measured in the three configurations are displayed in a single plot. The radial zone, where the fluctuation level drops, moves radially outwards from FQM to FMM configurations as the 5/5 island chain does. Thus, in the intermediate region between $\rho \sim 0.5 -0.7$, the density fluctuation level changes with the magnetic configuration, decreasing from FQM to FMM, 
which could be the reason for the observed increasing trend in plasma confinement reported in~\cite{Andreeva:2019,Andreeva:2021,Geiger:2021}.  
The change in the fluctuation level profile with the magnetic configuration can hardly be explained on the basis of local gradients in the plasma, as they remain almost unchanged (as shown in figure \ref{f:fig_1}). Thus, to investigate the possible relation between the reduction in the turbulence and the flow-shear developed associated to the magnetic island, the flow-shearing is calculated and is shown in figure \ref{f:fig_4}.b for the three magnetic configurations. A rather flat shearing rate profile is obtained in the FQM configuration (shown in blue), while a local maximum (in absolute value) is found at $\rho \sim 0.62 - 0.65$ in the FOM configuration (in green), which increases and moves towards $\rho \sim 0.7 - 0.72$ in the FMM configuration (in red). 
This result suggests an interpretation in terms of the combined effect of the flow-shearing and the radial spreading of turbulence: as the shearing increases, reduces the turbulence spreading from the plasma edge to the core.
The present experimental results resemble some experimental and simulation results~\cite{Wang:2007,Estrada:2011} where the key quantity to the control of turbulence spreading was found to be the flow-shearing rate.
Finally, in the FMM configuration the fluctuation level shows a slight modulation within the island region with a local minimum nearby the island O-point ($\rho \sim 0.83$). Concomitant with this reduction a maximum in the flow-shearing is also detected  (at $\rho \sim 0.86$). This observation will be further discussed in the next section.

\subsection{Heating power scan} \label{Heating power scan}

\begin{figure}[h]
 \center 
\includegraphics[width=0.55\columnwidth,trim= 20 20 0 0]{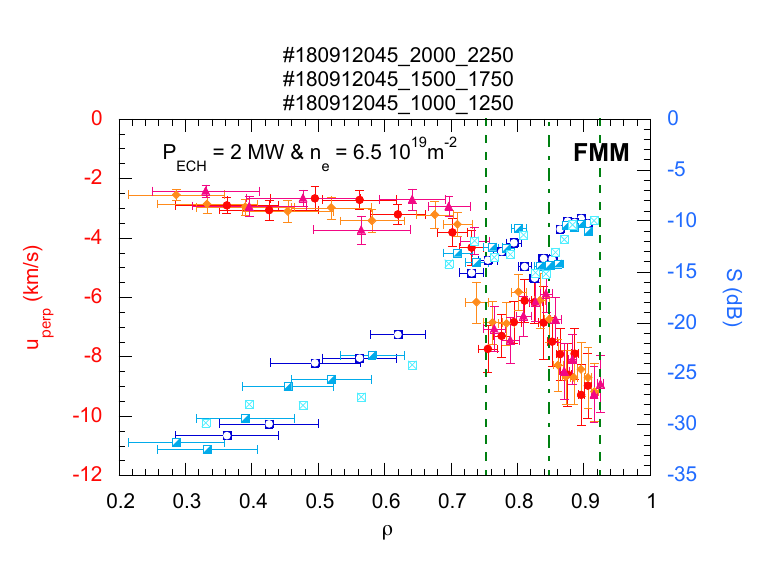}      
 \caption{Radial profiles of perpendicular plasma flow (in reddish colors) and density fluctuations (in blueish colors) measured by DR in the FMM magnetic configuration with $P_{ECH} = 2$ MW. Different symbols and color variations are used for the different time intervals. Vertical green lines indicate the DR measurement points closer to the island boundaries (dashed lines) and to the island O-point (dashed and dotted line) as obtained from the Poincar\'e plots.}
\label{f:fig_5}
\end{figure}

The impact of the ECH heating power has been studied in the FMM configuration by reducing the power up to $P_{ECH} = 2$ MW while keeping the plasma density constant at $n_e = 6.5$ $10^{19}$ m$^{-2}$. 
The radial profiles of the perpendicular plasma flow (in red) and density fluctuations (in blue) measured by DR are shown in figure \ref{f:fig_5}. 
As in previous cases, the DR measurement points closer to the island boundaries and to the island O-point have been identified in the Poincar\'e plot and represented in figure \ref{f:fig_5} as vertical green lines. 
It is worth mentioning that the island width estimated from the Poincar\'e plot ($\Delta \sim 4$ cm) does not change as it only depends on the specific magnetic configuration.
Some clear differences are found when these results are compared with those obtained in the plasma heated with $P_{ECH} = 4$ MW (shown in figure \ref{f:fig_2}.c).
In the 5/5 island region, the $W$-$shape$ pattern in the perpendicular plasma flow is also visible but in this case it shows an asymmetry, the flow at the outer island boundary being more intense than that measured at the inner island boundary.
Regarding the density fluctuations, a clear variation is visible within the island region with a minimum in the fluctuation level nearby the island O-point. As already pointed out, the same pattern is also visible in the plasma heated at $P_{ECH} = 4$ MW but less pronounced (figure \ref{f:fig_2}.c and red symbols in figure \ref{f:fig_4}). The notable minimum in the fluctuation level nearby the island O-point in the 2 MW case may be a consequence of the stronger flow-shear developed between the outer island boundary  and the island O-point.  
In addition, some differences are also found in the plasma bulk, $\rho < 0.7$, where both the plasma flow and the density fluctuations decrease as the heating power is reduced.

\subsection{Comparison with neoclassical $E\times B$ flow} \label{Comparison with neoclassical}

\begin{figure}[h]
 \center 
\includegraphics[width=0.49\columnwidth,trim= 20 20 0 0]{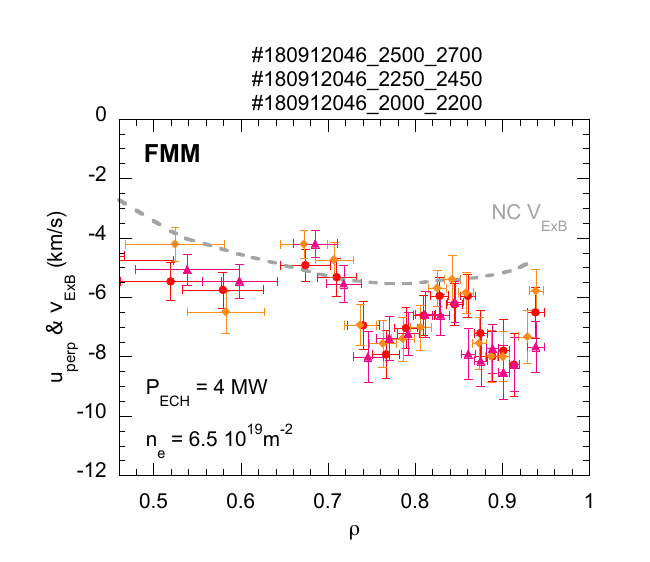}   
\includegraphics[width=0.49\columnwidth,trim= 20 20 0 0]{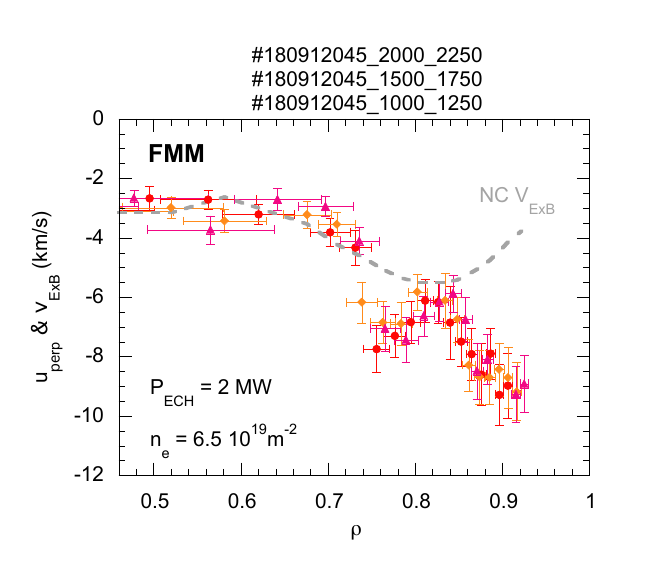}   
 \caption{Radial profiles of the perpendicular plasma flow measured by DR (symbols) and of the neoclassical $E\times B$ flow (grey broken line) calculated considering the neoclassical $E_r$ profile and the flux compression and magnetic field at the DR measurement region.}
\label{f:fig_6}
\end{figure}

In order to estimate the island contribution to the plasma flow, the perpendicular flow profile measured by DR has been compared with the neoclassical $E\times B$ flow. To that end, the neoclassical radial electric field, $E_r^{NC}$, is calculated using the neoclassical codes DKES (Drift Kinetic Equation Solver)~\cite{Hirshman:1986} and KNOSOS (KiNetic Orbit-averaging SOlver for Stellarators)~\cite{Velasco:2020,Velasco:2021}. These calculations consider nested flux surfaces and do not take into account the magnetic island. The local $E\times B$ flow is obtained considering the neoclassical $E_r^{NC}$, the flux compression at the DR measurement region, $| \nabla r|$, and the local magnetic field, B, as: $v_{E\times B}=E_r^{NC}| \nabla r| / B$.
Two scenarios, previously discussed, have been selected: both are in the FMM configuration with the same plasma density ($6.5$ $10^{19}$ m$^{-2}$) having different ECH heating power (4 and 2 MW). The result is shown in figure \ref{f:fig_6}: the symbols represent the perpendicular flow measured by DR and the broken grey line shows the neoclassical $E\times B$ flow. While outside the island region ($\rho < 0.75$), the perpendicular flow measured by DR and the neoclassical $E\times B$ flow show a rather good agreement, in the island zone the profiles deviate from each other in the two plasma scenarios. The difference between the flow profiles may be considered as an estimation of the island contribution to the plasma flow which is maximum at the island boundaries and close to zero at the island O-point. In the plasma heated with $P_{ECH}=4$ MW, the island contribution to the flow is rather similar at the inner and outer island boundaries; in the $P_{ECH}=2$ MW case, however, the island contribution to the flow is twice as large at the outer island boundary, closer to the plasma edge, than at the inner island boundary.

\section{Summary and Discussion} \label{Summary and Discussion}

The effect of the 5/5 magnetic island on plasma flow and turbulence has been experimentally investigated in the stellarator W7-X using a Doppler reflectometry diagnostic.
The plasma flow profile measured in the FMM magnetic configuration (edge rotational transform equal to 1.05) shows a remarkable $W$-$shape$ pattern in the radial range $\rho \sim 0.70 - 0.95$. Poincar\'e plots obtained from field line tracing calculations reveal a rather good matching between the location of the $W$-$shape$ flow pattern and the 5/5 magnetic island region. 
These results show that the island contribution to the plasma flow, estimated as the difference between the experimental flow profile and the neoclassical $E\times B$ flow profile, is maximum at the island boundaries and close to zero at the island O-point.
This result resembles the flow profile measured across the 1/1 magnetic island in  LHD plasmas when the island width does not exceed a critical value (15\% - 20\% of the minor radius)~\cite{Ida:2001}. 
Similarly, gyrokinetic simulations show that the island contribution to the flow is localised at the island boundaries when the island width is below a threshold~\cite{Banon_Navarro:2017}. These simulations also show that above this threshold, the flow reverses on one side of the island and a vortex-like flow develops around the island O-point. The onset of the vortex-like flow occurs simultaneously with the flattening of the temperature profile. This vortex-like flow, found in LHD~\cite{Ida:2001} and TJ-II~\cite{Estrada:2016} for large islands, is not observed in the present W7-X experiments which may indicate that the island width does not exceed the threshold for the onset of the vortex-like flow and temperature flattening.

Two radial zones can be distinguished in the density fluctuation profile, an external region with high fluctuation level, and an internal one towards the plasma core where the fluctuation level drops. The radial position where the drop in the fluctuations is detected moves as the magnetic island position changes. This result, which can hardly be explained on the basis of local gradients in the plasma, could be linked to the increase in the flow-shearing detected nearby the inner island boundary which may reduce the turbulence spreading from the plasma edge towards the core.

The results reported in the present work also show that the island contribution to the flow becomes asymmetric in plasmas heated with lower ECH power (2 MW instead of 4 MW for the same plasma density $n_e \sim 6.5$ $10^{19}$ m$^{-2}$). The island contribution to the flow in this case is twice as large at the outer island boundary than at the inner island boundary. 
In this case, the density fluctuation level shows a pronounced minimum nearby the island O-point.
These results resemble those found in J-TEXT~\cite{Zhao:2015}, where the flow is enhanced at the outer island boundary when the magnetic island approaches the LCFS, and the density fluctuations drop inside the magnetic island. 
Such an asymmetric flow is also found in gyrokinetic simulations when a radially asymmetric magnetic island is considered~\cite{Banon_Navarro:2017}. 
Stronger flow and flow shear at the outer island boundary, as found in the present experiments, are obtained for a magnetic island with higher curvature at the outer island boundary. 
It has to be noted that, in these experiments, we do not have any experimental evidence of asymmetric magnetic islands as found in other devices~\cite{Snape:2012,Bardoczi:2016b}. Moreover, the reason why the island geometry could be different in the 2 MW and in the 4 MW heated plasmas is unknown. We point to the island geometry effect as a possible explanation due to the similarities between our measurements and the GK simulation results.
In the simulations, the authors also stress that the asymmetric flow shear reduces turbulence penetration into the island on the side where the shear is increased. This could explain the minimum in the density fluctuation level measured nearby the island O-point, which is more pronounced in the plasma heated with 2 MW showing a higher flow shear. 
This minimum, however, could be also consequence of a flattening in the plasma profiles at the O-point, which, as already mentioned, cannot be ruled out in the present experiments.

 In summary, magnetic configuration scans performed at W7-X show modifications of the plasma flow and density fluctuations associated to the 5/5 magnetic island in a limiter magnetic configuration with $\iota = 1$ located at the plasma edge but inside the confinement region (at  $\rho \sim 0.8$). 
The 5/5 magnetic island modifies the local flow, increases the flow-shear and produces a reduction in the density fluctuation level.
The island contribution to the plasma flow is maximum at the island boundaries and close to zero at the island O-point. This contribution becomes asymmetric at low ECH heating power, with stronger flow and flow shear at the outer island boundary.
Density fluctuations show a local minimum nearby the island O-point, more pronounced in the case with higher flow shear. Besides, the radial region, where the density fluctuations drop, shifts with the radial position of the island. 
W7-X results show similarities with results found in other devices and in GK simulations. The comparison with the latter suggest that in the present experiments, the 5/5 island width does not exceed the threshold found in the simulations for the onset of the vortex-like flow and temperature flattening. Besides, the asymmetric island contribution to the flow found at low ECH power may be explained by GK simulation results for radially asymmetric magnetic island with higher curvature at the outer island boundary.

\section*{Acknowledgements}

\small{The authors acknowledge the entire W7-X team for their support.
This work has been partially funded by the Spanish Ministry of Science and Innovation under contract numbers FIS2017-88892-P and PGC2018-095307-B-100. 
This work has been carried out within the framework of the EUROfusion Consortium and has received funding from the Euratom research and training programme 2014-2018 and 2019-2020 under grant agreement No 633053. The views and opinions expressed herein do not necessarily reflect those of the European Commission.}

\section*{References} 

\bibliographystyle{prsty_copia}

\bibliography{Bibtex_database_copia}

\end{document}